\documentclass[sn-mathphys]{sn-jnl}

\usepackage{latexsym}
\RequirePackage{graphicx}
\RequirePackage{mathptmx}      
\usepackage[bottom]{footmisc}
\usepackage{float}

\usepackage{amsmath}
\usepackage{subfigure}

\jyear{2021}%

\theoremstyle{thmstyleone}%
\theoremstyle{thmstyletwo}%

\theoremstyle{thmstylethree}%

\begin{document}

\title[Beam steering with quasi-mosaic bent silicon single crystals]{Beam steering with quasi-mosaic bent silicon single crystals
}
\subtitle{Computer simulations for 855 MeV and 6.3 GeV electrons and comparison with experiments}

\author{\fnm{H.} \sur{Backe}}\email{backe@uni-mainz.de}
\affil{\orgdiv{Institute for Nuclear Physics}, \orgname{Johannes Gutenberg-University} \city{Mainz}, \postcode{D-55128},  \country{Germany}}


\abstract{Monte Carlo simulations have been performed for 855 MeV and 6.3 GeV electrons channeling in silicon single crystals at circular bent (111) planes. The aim was to identify critical experimental parameters which effect the volume-deflection and volume-capture characteristics. To these belongs the angular alignment of the crystal with respect to the nominal beam direction. The continuum potential picture has been utilized. The simulation results were compared with experiments. It turns out that the assumption of an anticlastic bending of the crystal, bent on the principle of the quasi-mosaic effect, is not required to reproduce the experimental observations.}

\maketitle

\section{Introduction}
\label{intro}
The understanding of the de-channeling process in bent single crystals is of utmost importance, not only for particle steering in high energy physics but also for the construction of compact radiation sources in the MeV range and beyond, for an overview see e.g. Korol et al. \cite{KorS14}. In all cases bent crystals are required the production of which are based on different principles. The quasi-mosaic effect has been proven as beneficial to achieve a curvature of certain plains of single crystals which otherwise would be flat \cite{CamG15}. A number of experiments utilized this effect to steer charged particles at (111) planes of silicon single crystals. For the current work the experimental publications \cite{MazB14,WieM15,WisU16} are of particular importance in which the results of deflection studies for 855 MeV and 6.3 GeV electrons have been described.

It is well known that at bending employing the quasi-mosaic effect a (parasitic) anticlastic bending is connected which must be minimized \cite{CamG15,GuiM09}. If it is significant, the beam steering pattern may alter as function of the anticlastic bending radius, the beam divergence and spot size, as well as its lateral displacement with respect to the center of the spherical calotte. Such effects were recently investigated on the basis of simulation calculations with the MBN explorer software package by Sushko et al. \cite{SusK21} suggesting that the beam profiles measured in the SLAC experiment \cite{WieM15,WisU16} cannot be understood without the introduction of an anticlastic bending.

In this contribution some key experimental results obtained for silicon single crystals were reinvestigated by means of Monte Carlo simulations with a computer code utilizing the well known continuum potential picture of Lindhard \cite{Lin65}. The latter has been exploited by a number of authors, for an overview see, e.g., Korol et al. \cite[Chapter 2]{KorS21}. In particular the DYNECHARM++ code of Bagli and Guidi \cite{BagG13}, and CRYSTALRAD of Sytov et al. \cite{SytT19} should be mentioned. The continuum potential picture has the advantage that channeling can be classically understood for ultra-relativistic particles in a rather intuitive manner. The current work is based on the paper of Backe \cite{Bac22} in which details of the underlying formalism are described on the example of channeling of 855 MeV electron in (110) planes of diamond.

The basic ingredients of the model are summarized in the following section \ref{basics}. Section \ref{Calculations} is devoted to the Monte Carlo simulations of the beam profiles. The results are compared with the experimental observations of beam profile measurements \cite{MazB14,WieM15,WisU16}, and discussed in section \ref{discussion} with regard to a possible anticlastic bending. The paper closes in section \ref{conclusions} with conclusions.  For the purpose of a check of the reliability of the code, in Appendix \ref{appendix B} simulation calculations are presented for the transition rate as function of the penetration depth and compared with experimental results.

\section{Continuum potential and scattering distributions} \label{basics}
\begin{figure}[tb]
\centering
    \includegraphics[angle=0,scale=0.45,clip]{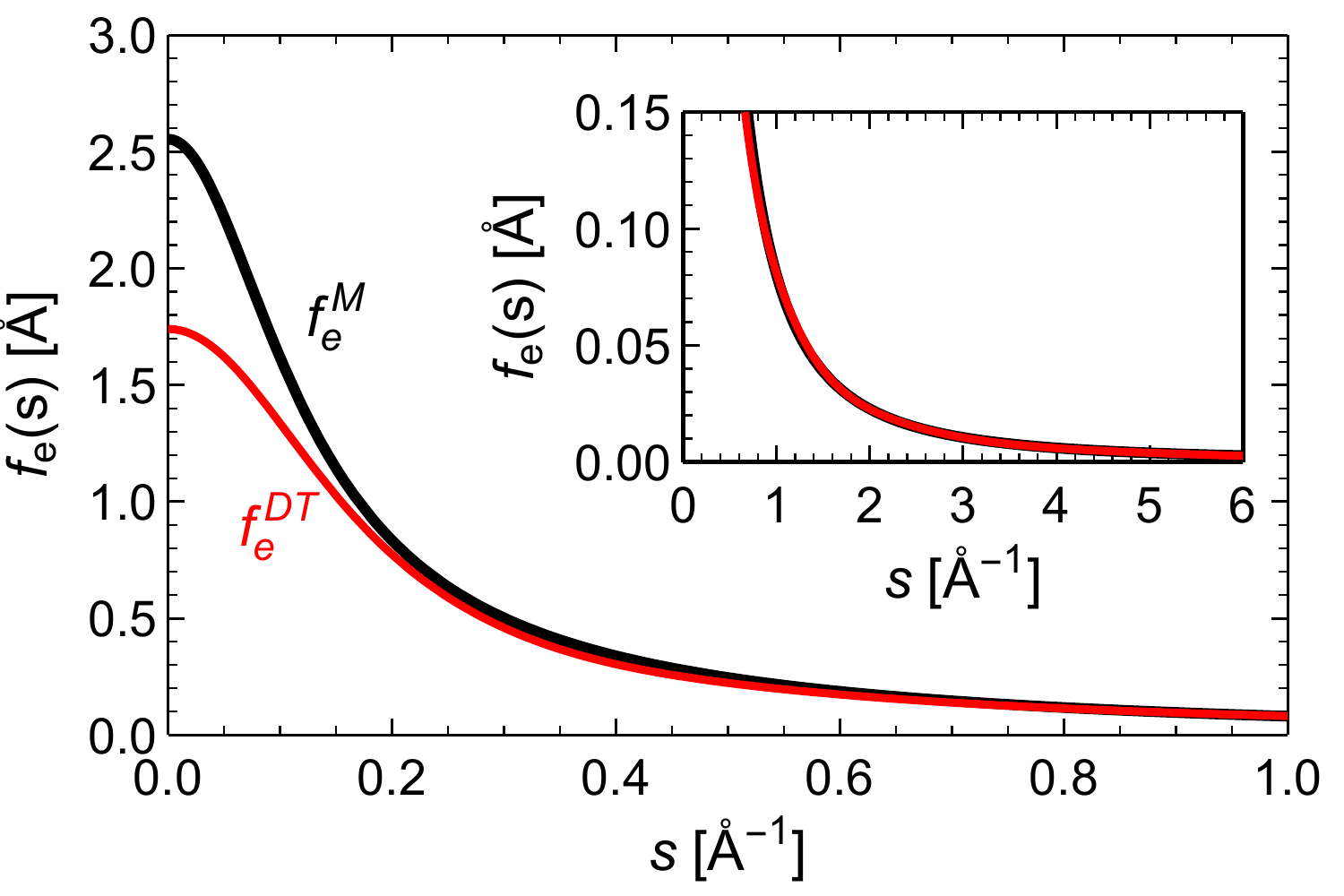}
\caption[]{Electron scattering factors for atomic silicon. The black full curve $f_e^{M}$ represents the Moli\`{e}re approximation \cite{Mol47} with the original parameter set, the red one $f_e^{DT}$ that of Doyle and Turner \cite{DoyT67} in the six-parameter representation according to Ref. \cite{ChoU99}.} \label{ScatteringFactors}
\end{figure}

For the calculation of the channeling potential, as well as the angular distribution of the scattered electrons, the Moli\`{e}re representation for the electronic scattering factors \cite{Mol47} was used, however, with the modified parameters for silicon
\begin{eqnarray}\label{MoliereParamMod}
\alpha &=& \{0.514906,~~0~~,~0.485094\}\nonumber \\
\beta &=& \{2.113681,~1.2,~0.41510\}/a_{TF}.
\end{eqnarray}
Here are $a_{TF}=0.8853 ~a_0 ~Z^{-1/3}$ the Thomas-Fermi screening factor with $a_0$ the Bohr radius, and $Z$=14 for silicon. This parameter set approximates the six-parameter Doyle-Turner representation quoted by Chouffani and \"{U}berall \cite{ChoU99}, see Fig. \ref{ScatteringFactors}, to better than 8\% in the full range of $0 \leq s/{\AA} \leq 6$. The quantity $s$ is related to the momentum transfer by $q=2p v/(\hbar c) \sin(\vartheta/2) = 4\pi s$, with $p$ the momentum of the particle, $v$ its velocity, $c$ the speed of light, and $\vartheta$ the scattering angle.

\begin{figure}[b]
\centering
    \includegraphics[angle=0,scale=0.55,clip]{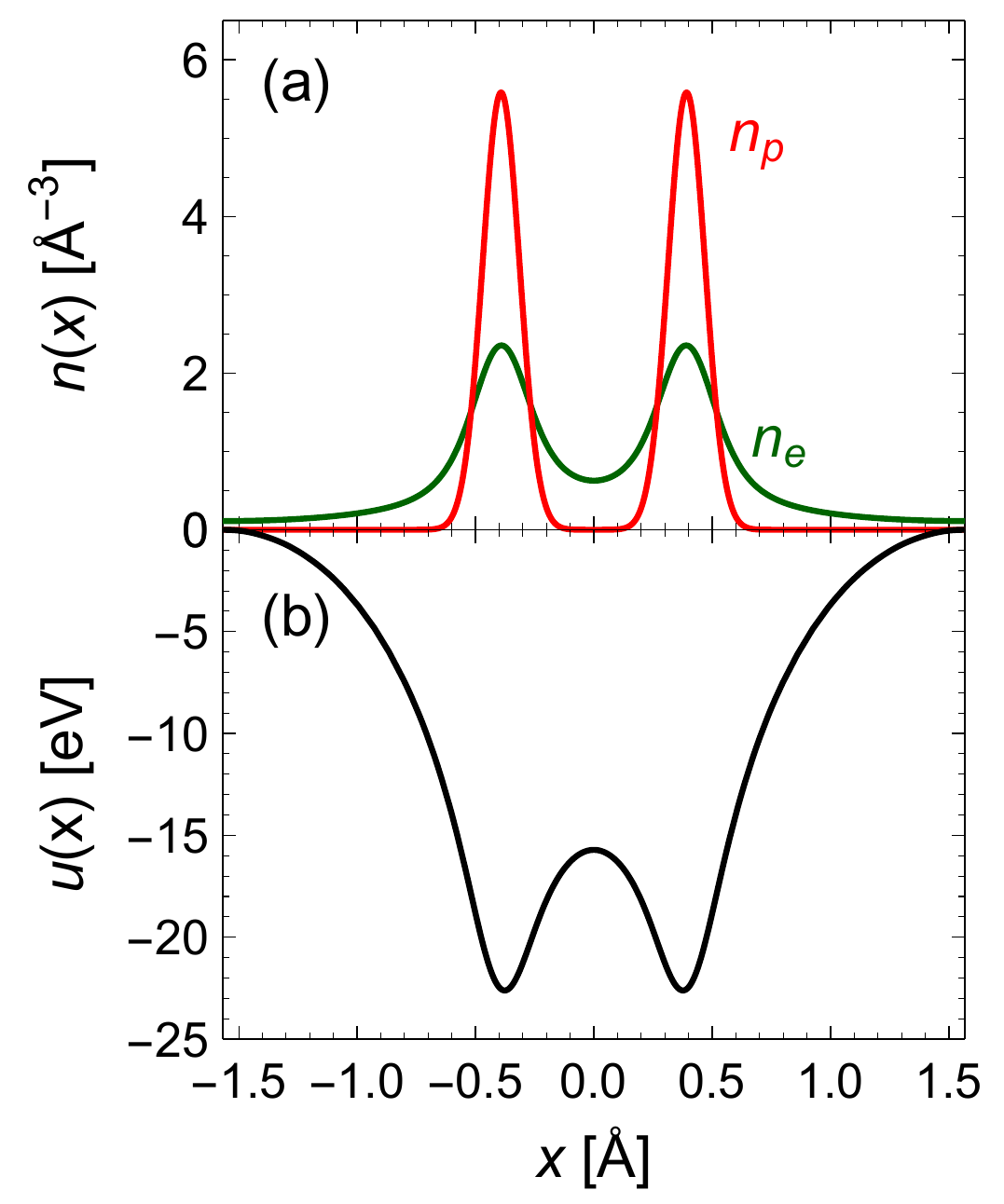}
\caption[]{(a) Electron charge density $n_e(x)$ and positive charge density $n_p(x)$ as function of the distance coordinate $x$ across the (111) planes of  a silicon single crystal. The inter-planar distance amounts to $d_p~=~3.136~$ {\AA}. The integral $(1/d_p)\int_{-d_p/2}^{d_p/2} n_e(x) dx = Z\cdot (8/a^3)$ = 0.699/${\AA}^3$ represents the mean electron density with $a$ = 5.431 ${\AA}$ the lattice constant \cite{website:ioffe}. (b) Potential, the depth amounts to  $u_0$ = -22.615 eV.} \label{potential}
\end{figure}
The potential $u(x)$ has been calculated according to the textbook of Baier et al. \cite[chapter 9.1]{BaiK98}. The result is shown in Fig. \ref{potential} together with the positive and negative charge densities $n_p(x)$ and $n_{el}(x)$, respectively, across the channel. The latter has been derived from the plane potential with the aid of the one dimensional Poisson equation.

\begin{figure}[b]
\vspace*{-0.5cm}
\centering
    \subfigure{
    \includegraphics[angle=0,scale=0.6,clip]{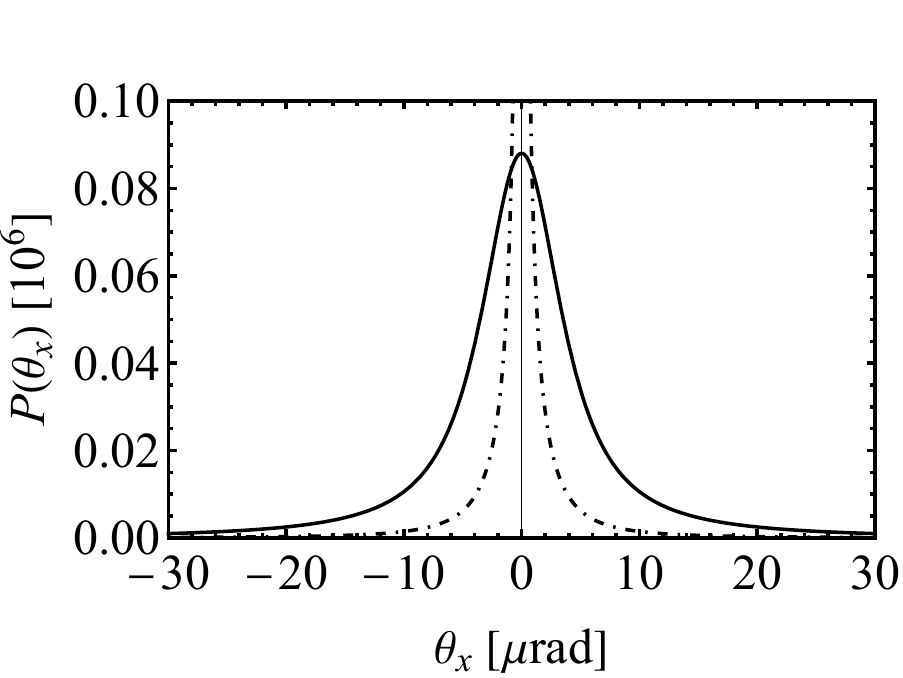}}
    \subfigure{
    \includegraphics[angle=0,scale=0.6,clip]{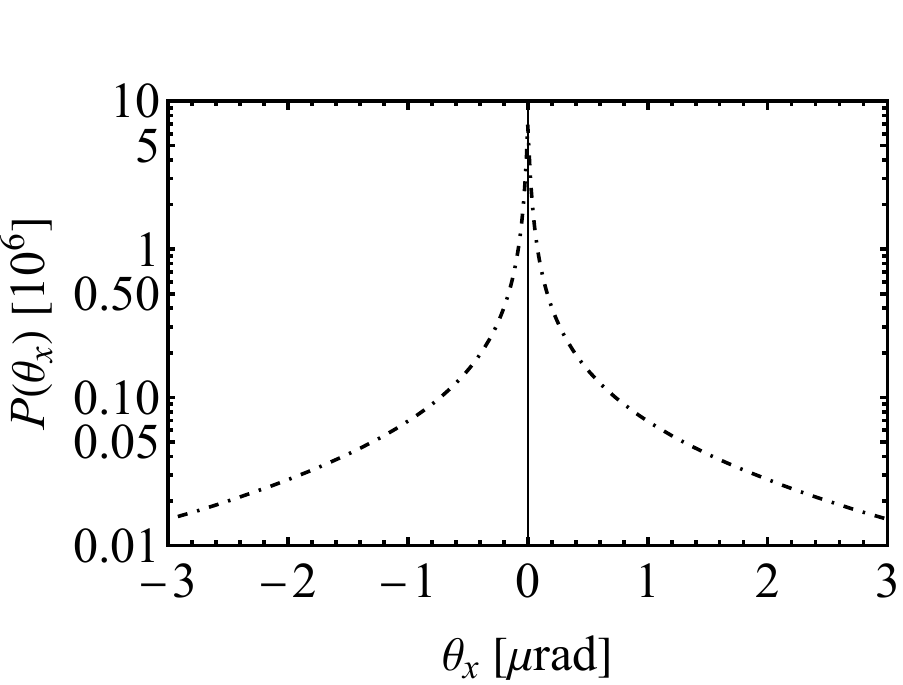}}
\caption[]{Normalized electron-atom (full) and electron-electron (dot-dashed) scattering distributions for 855 MeV electrons. In the right panel a cutout is shown in a logarithmic representation. The FWHM amount to 8.03 $\mu$rad (atomic) and 0.033 $\mu$rad (electronic). Both distributions have long tails taken into account in the numerical simulation up to $\pm$ 0.0345 rad. The root mean squared scattering angles amount to $<\theta_x^2>^{1/2}$ = 23.3 $\mu$rad (atomic), and 7.42 $\mu$rad (electronic). The total scattering cross-sections are $\sigma_{tot}^{(at)}$ = 89.38$\cdot\ 10^{-4}~\AA^2$ and $\sigma_{tot}^{(el)}$ = 5.158 $\cdot 10^{-4}~\AA^2$, the mean transverse energy gains $<\Delta E_\bot/\Delta z>_{at}$ = 1.040 eV/$\mu$m and $<\Delta E_\bot/\Delta z>_{el}$ = 0.0853 eV/$\mu$m, and the mean number of collisions 4.46/$\mu$m (atomic) and 3.606/$\mu$m (electronic).} \label{ProbabilityWnorm855}
\end{figure}

\begin{figure}[tbh]
\vspace*{-0.5cm}
\centering
    \subfigure{
    \includegraphics[angle=0,scale=0.60,clip]{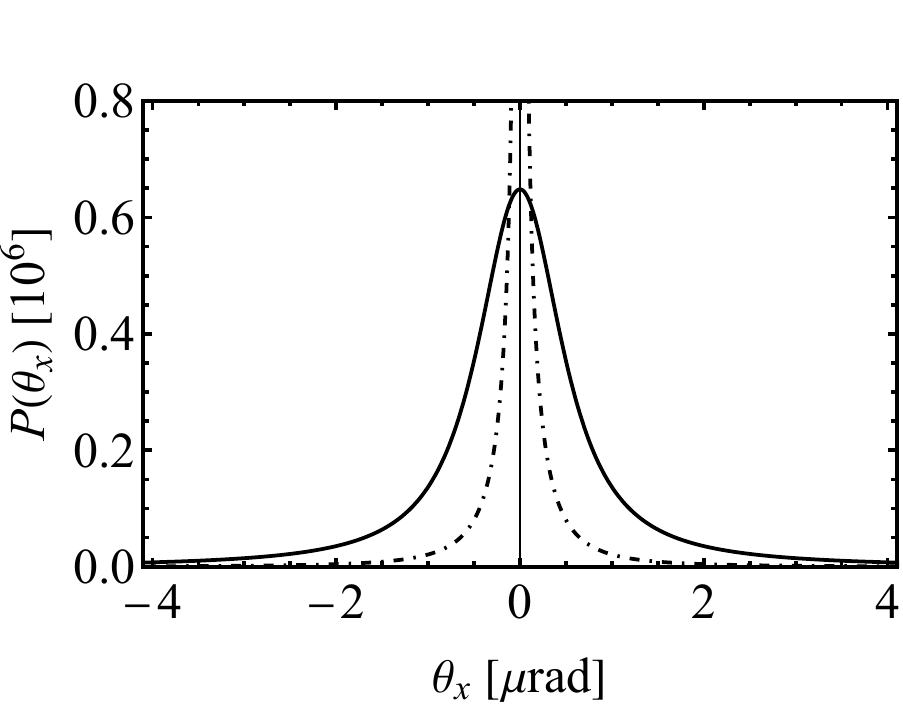}}
    \subfigure{
    \includegraphics[angle=0,scale=0.6,clip]{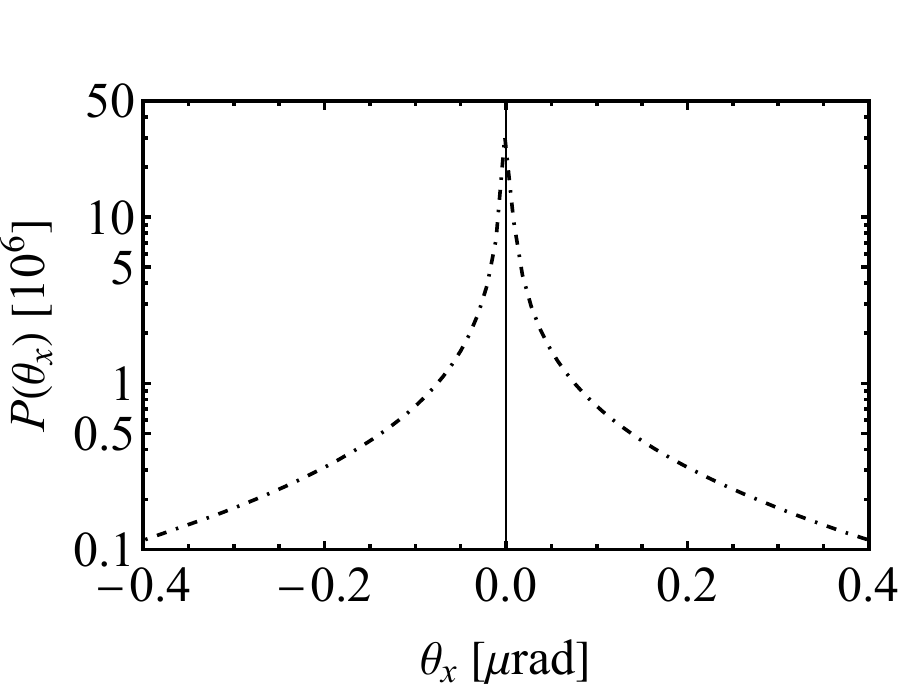}}
\caption[]{Normalized electron-atom and electron-electron (dot-dashed) scattering distributions for 6.3 GeV electrons. In the right panel a cutout is shown in a logarithmic representation. The FWHM amount to 1.09 $\mu$rad (atomic) and 0.00868 $\mu$rad (electronic). Both distributions have long tails taken into account in the numerical simulation up to $\pm$ 0.012 rad. The root mean squared scattering angles amount to $<\theta_x^2>^{1/2}$ = 3.49 $\mu$rad (atomic), and 1.08 $\mu$rad (electronic). The total scattering cross-sections are $\sigma_{tot}^{(at)}$ = 89.38$\cdot\ 10^{-4}~\AA^2$ and $\sigma_{tot}^{(el)}$ = 5.158 $\cdot 10^{-4}~\AA^2$, the mean transverse energy gains $<\Delta E_\bot/\Delta z>_{at}$ = 0.171 eV/$\mu$m and $<\Delta E_\bot/\Delta z>_{el}$ = 0.0130 eV/$\mu$m, the mean number of collisions 4.46/$\mu$m (atomic) and 3.606/$\mu$m (electronic).} \label{ProbabilityWnorm6300}
\end{figure}

The normalized scattering distribution functions for electron-atom interactions at energies of 855 MeV and 6.3 GeV, respectively, are
\begin{eqnarray} \label{normalScattProb855}
P_{855}^{(at)}(\theta_x) =\Big(\frac{1.66711\cdot 10^{-10} + \theta_x^2}{(2.42622\cdot 10^{-11} + \theta_x^2)^{3/2}}
\nonumber\\
& &\hspace{-2.5 cm}
-\frac{4.68581\cdot 10^{-10} + \theta_x^2}{(6.29077\cdot 10^{-10} + \theta_x^2)^{3/2}}\Big)/15.5081,
\end{eqnarray}
\begin{eqnarray} \label{normalScattProb6300}
P_{6300}^{(at)}(\theta_x) =\Big(\frac{3.07372\cdot10^{-12} +\theta_x^2}{(4.47332\cdot 10^{-13} + \theta_x^2)^{3/2}}
\nonumber\\
& &\hspace{-2.5 cm}
-\frac{8.63941\cdot10^{-12} + \theta_x^2}{(1.15985\cdot 10^{-11} + \theta_x^2)^{3/2}}\Big)/15.5081
\end{eqnarray}
with $\theta_x$ the projected scattering angle. The numerical parameters are related to the Moli\`{e}re parameters of Eq. (\ref{MoliereParamMod}) in a rather involved manner. The distribution functions are depicted in Figs. \ref{ProbabilityWnorm855} and \ref{ProbabilityWnorm6300}, full curves.

The scattering distributions for electron-electron collisions have also been calculated according to \cite{Bac22}, for some details see Appendix \ref{appendix A}. The results are included in Figs. \ref{ProbabilityWnorm855} and  \ref{ProbabilityWnorm6300} as dot-dashed lines. These distributions can be approximated by the heuristic function\footnote{For electrons with an energy of 855 MeV the parameters are  $A_{855}$ = 3.5 $\cdot 10^{-12}$, $a_{855} =4.5\cdot 10^{-19} $,  $b_{855} =4.0 \cdot10^{-11} $ and $c_{855} =1.0 \cdot10^{-5} $, and for the energy of 6.3 GeV $A_{6300}$ = 6.7 $\cdot 10^{-14}$, $a_{6300} =2.0\cdot 10^{-21} $,  $b_{6300} =7.2 \cdot10^{-13} $ and $c_{6300} =1.5 \cdot10^{-6}$. The approximations are better than about  $\pm$ 10 \%  in the interval  0.1 $\mu$rad  $< \lvert\theta_x \rvert<$ 10 mrad and 0.1 $\mu$rad  $< \lvert\theta_x \rvert <$ 5 mrad for 855MeV and 6300 MeV, respectively. Below the lower and above the upper limits the approximations are significantly worse.}
\begin{eqnarray}\label{Pel855MeVapprox}
P^{(el)}(\theta_x)=\frac{A}{\lvert\theta_x^3 \rvert + c \cdot \theta_x ^2 + b\cdot \lvert\theta_x\rvert + a}.
\end{eqnarray}

\section{Simulation calculations} \label{Calculations}
Based on the scattering distributions derived in the last section, simulation calculation have been performed for (111) planar channeling of 855 and 6.3 GeV electrons in silicon single crystals. Details of the simulation procedure are described in the paper \cite{Bac22}.
For the former energy the crystal had a thickness of 30.5 $\mu$m and a bending radius of $R$ = 33.5 mm, for the latter 60 $\mu$m and $R$ = 150 mm. In both cases experimental results are available \cite{MazB14,WisU16} with which the simulation calculations will be compared. The reliability of the formalism was checked with results for the rate distribution as function of the penetration depth from which de-channeling lengths were derived for a comparison with literature, see appendix \ref{appendix B}.
\subsection{Sensitive parameters for the shape of beam profiles} \label{sensitive parameters}
\begin{figure}[t]
\centering
    \subfigure{
  \includegraphics[angle=0,scale=0.45,clip]{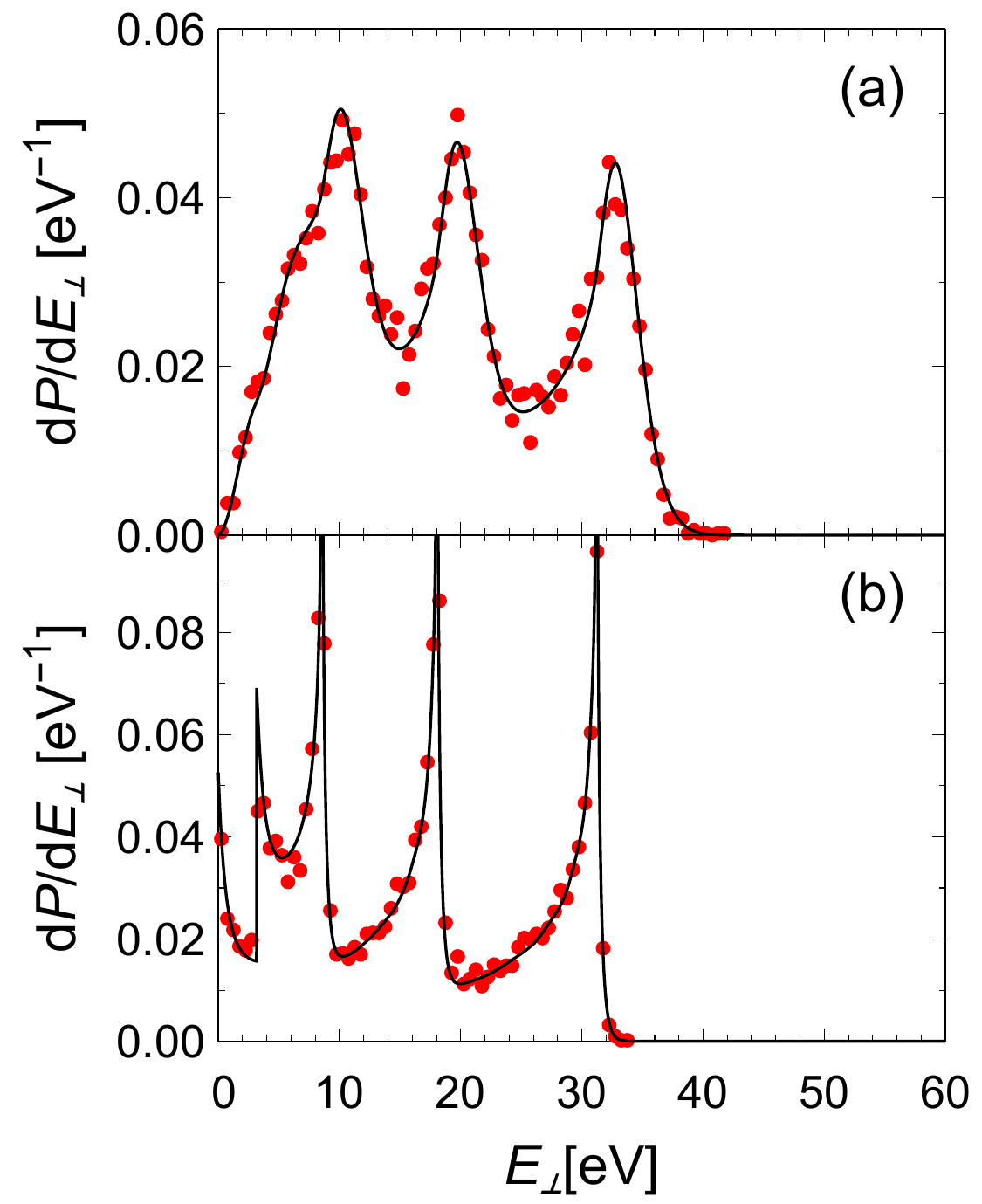}}
  \subfigure{
  \includegraphics[angle=0,scale=0.45,clip]{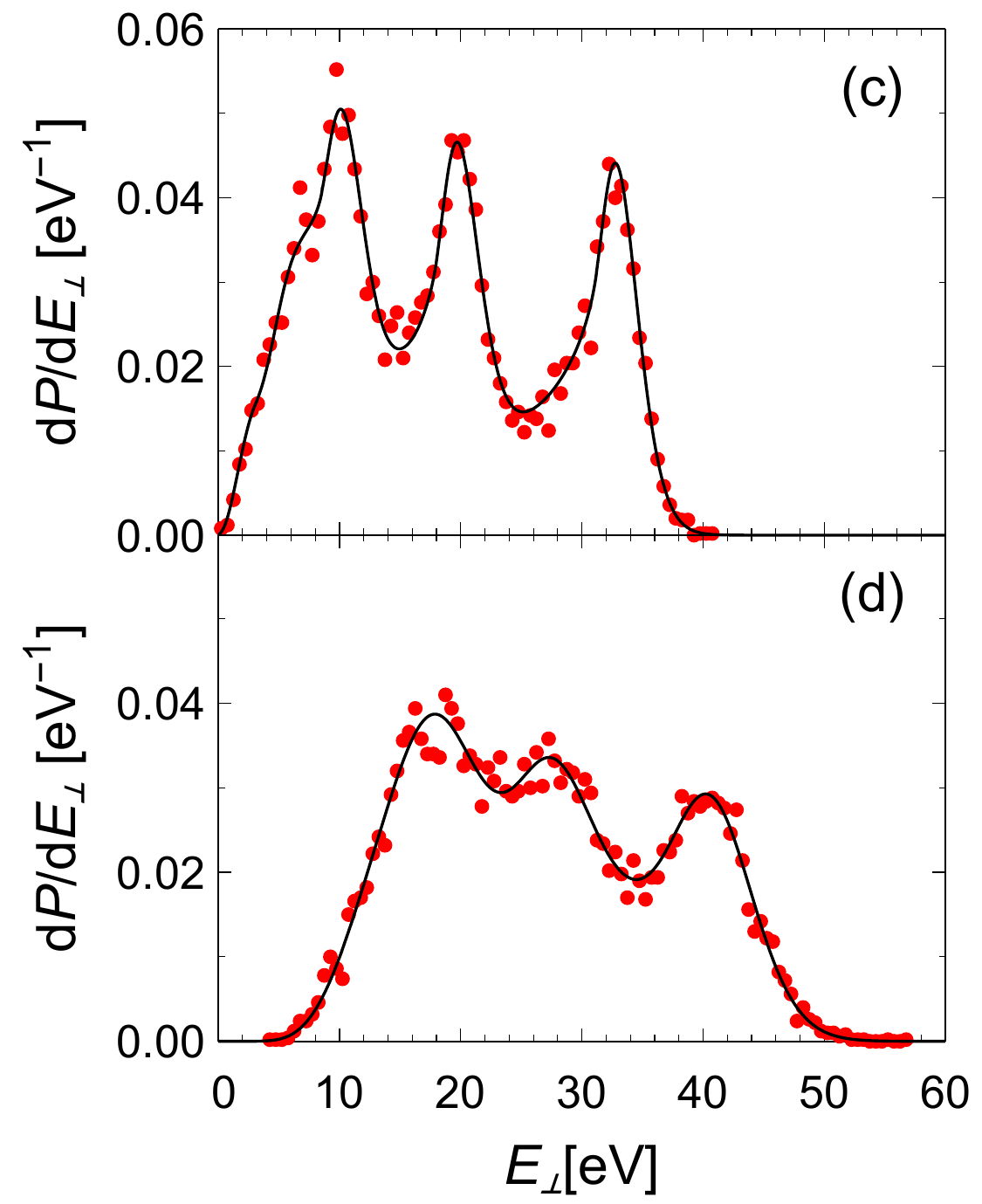}}
\caption[]{Initial probability distributions as function of the transverse energy $E_\bot$ for electrons of 6.3 GeV energy entering the silicon crystal with bending radius of R = 150 mm with angles of $\psi_{0}$ = (a) 30, (b) 0, (c) -30, and (d) -60 $\mu$rad. The thin black curve represents analytical calculations \cite[Eq. (15)]{BacL08}, the red dots are simulation calculations with 10,000 trials. In $E_\bot = (pv/2) (\vartheta_x-\psi_{0})^2 - U(x)$, $U(x) = u(x) -(pv/R)\cdot x-u_1$ is the deformed potential with $u_1$ the minimum of $U(x)$ in the channel, position $x$ and entrance angle $\vartheta_x$ are random variables ($\sigma '_{\vartheta x}$ = 8.0 $\mu$rad). Effects of a possible anticlastic bending were excluded.} \label{InitialCalcSim}
\end{figure}

Of crucial importance for the shape of the beam profiles is the initial occupation probability $d P/d E_\bot$as function of the transverse energy $E_\bot$. As shown in Fig. \ref{InitialCalcSim} it strongly depends on the alignment of the crystal with respect to the beam direction, i.e., on the entrance angle $\psi_{0}$ into the crystal. A uniform distribution of the electron density across the transverse $x$ coordinate was assumed, and for the angular distribution a Gaussian with standard deviations $\sigma '_{\vartheta x}$ = 8.0 $\mu$rad, as conjectured from the information given by Wienands et al. \cite{WieM15}.

In addition, sensitive parameters are the anticlastic bending radius, and a misalignment of the electron beam with respect to the spherical calotte representing the anticlastic deformation. Nearly no experimental information was provided in the publications except that the anticlastic bending was minimized at the production of the crystals. Significant effects on the beam profile for the SLAC experiment are expected for an anticlastic bending radius less than 10 m, assuming for spot size and divergence the upper limits of 150 $\mu$m and 10 $\mu$rad, respectively.

Initially, an anticlastic bending will be neglected in the analysis. If the experimental observations can not be reproduced, in a second step assumptions on the anticlastic bending radius, on the beam spot (size, divergence, displacement) can be made in order to improve the simulation results.

\subsection{Beam profiles for the MAMI experiment at 855 MeV} \label{profiles 855 MeV}
\begin{figure}[t]
      \subfigure{
      \includegraphics[angle=0,scale=0.188,clip]{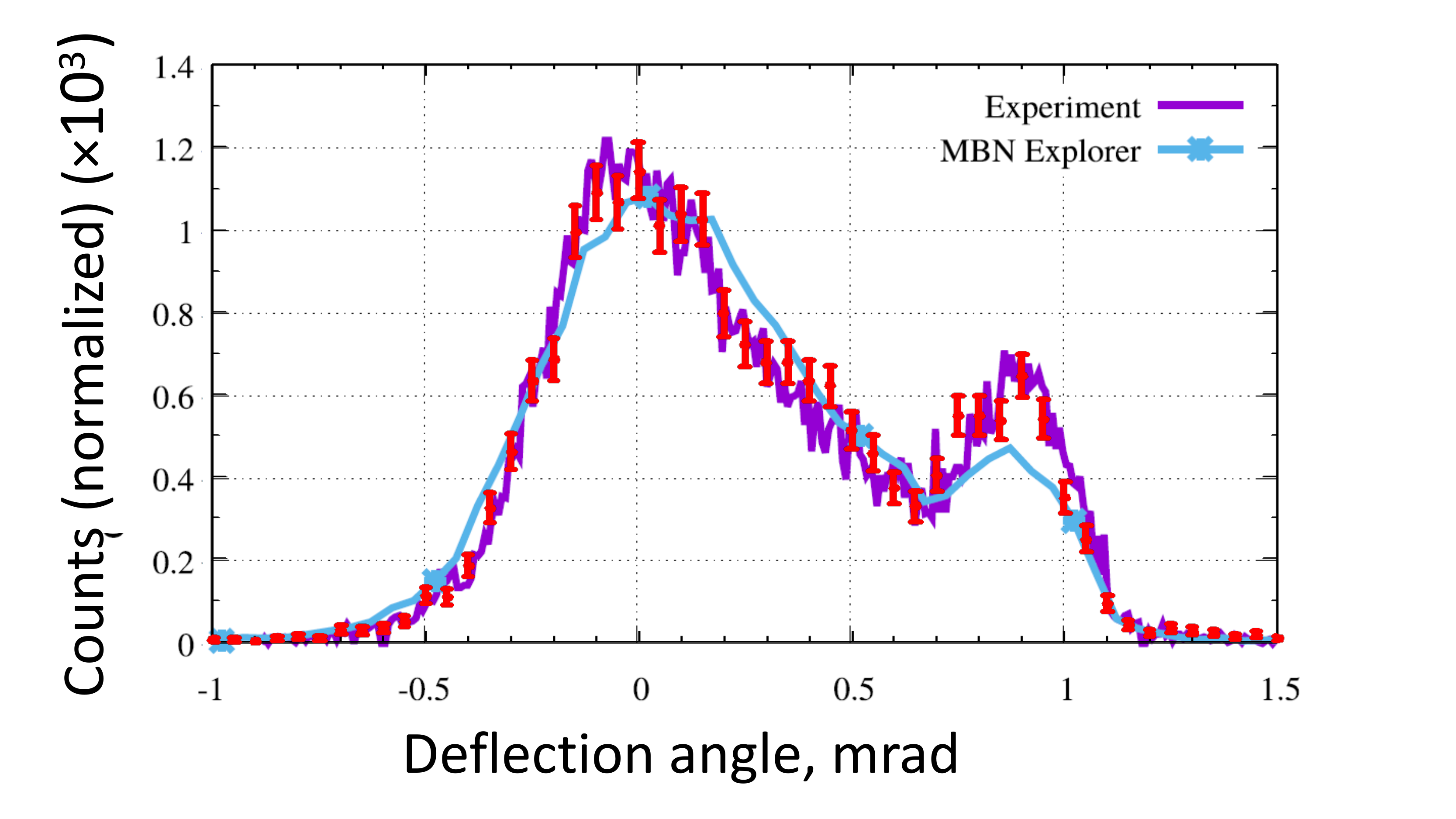}}
\hspace*{-0.7cm}
     \subfigure{
    \includegraphics[angle=0,scale=0.188,clip]{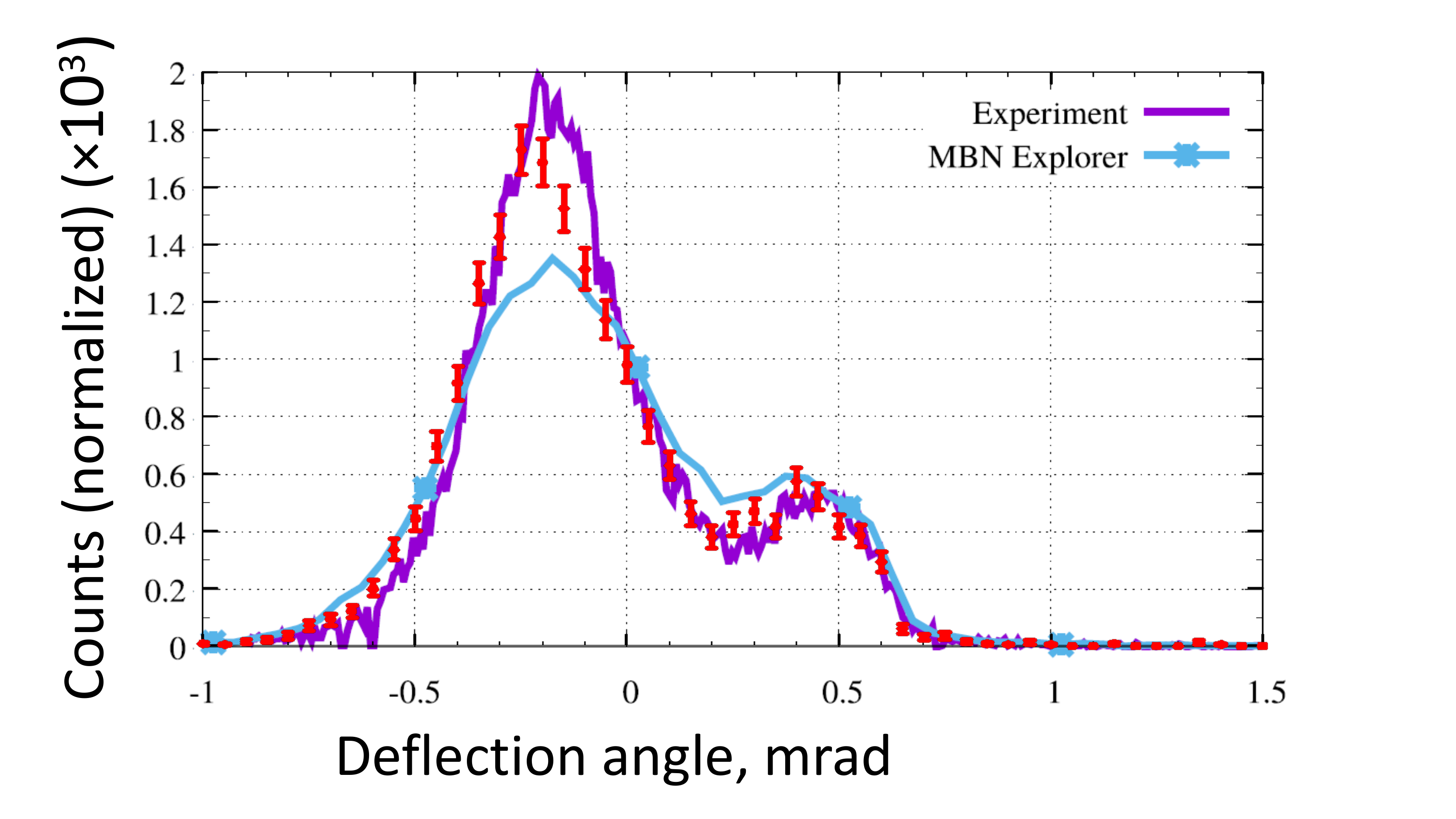}}
\caption[]{
Beam profiles for a 30.5 $\mu$m thick silicon single crystal with a bending radius of 33.5 mm at an electron beam energy of 855 MeV, entrance angles $\psi_0$ = 0 $\mu$rad (left panel), and $\psi_0$ = 450 $\mu$rad (right panel). For the simulation, red error bars (this work), a beam spot size of 29.7  $\mu$m (1 $\sigma$), and a beam divergence of 12.7  $\mu$rad (1 $\sigma$) were assumed, neglecting anticlastic bending. Light blue curves show simulations with the MBN explorer software package taken from Haurylavets et al. \cite[Fig. 9 and 11]{HaL22}. Experimental data, violet curve, according to Mazzolari et al. \cite{MazB14}.} \label{beamProfilesBent855MeVPsim400murad}
\end{figure}

The simulated beam profiles for 855 MeV electrons steered in (111) planes of the silicon single crystal are shown in Fig. \ref{beamProfilesBent855MeVPsim400murad}, together with experimental results taken from Mazzolari et al. \cite{MazB14}. A good overall agrement can be stated. In particular, the deflection peaks due to channeling, as well as the volume reflection shift in opposite direction for the tilted crystal are well reproduced. Merely, the intensity of the volume reflected peak seems slightly lower than the measurement, and the shift a little bit larger. It should be stressed that no change has been observed in the simulation calculations within the statistical errors if an anticlastic bending radius of 3.66 m \cite{GuiM09} was assumed.

\begin{figure}[tb]
    \subfigure {\includegraphics[angle=0,scale=0.47,clip]{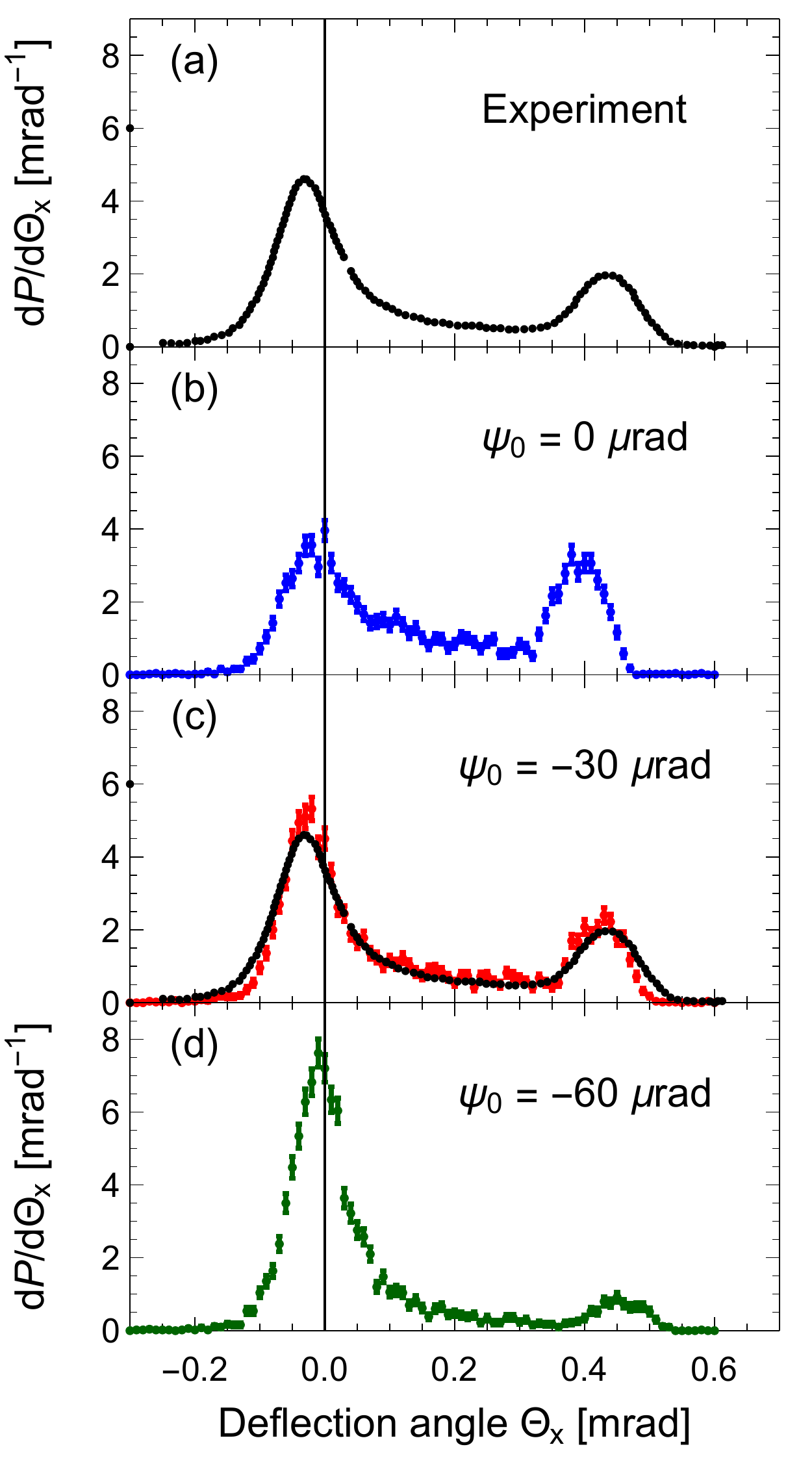}}
    \subfigure {\includegraphics[angle=0,scale=0.47,clip]{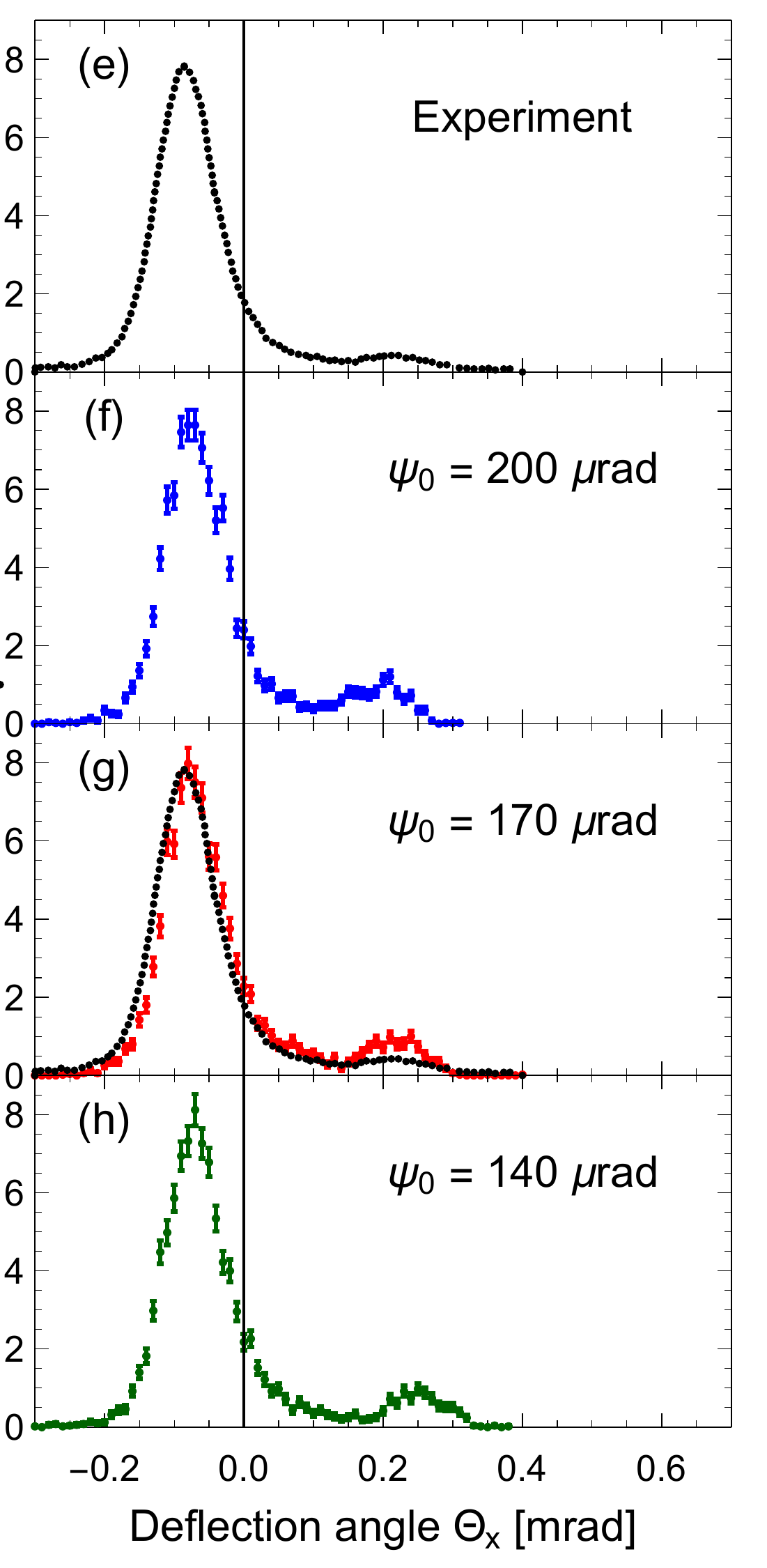}}
\caption[]{Beam profiles for a 60 $\mu$m thick silicon single crystal with a bending radius of 150 mm at an electron energy of 6.3 GeV for entrance angles $\psi_{0}$ as indicated. A beam spot size of 130 $\mu$m (1 $\sigma$), and a beam divergence of 8 $\mu$rad (1 $\sigma$) were assumed, neglecting anticlastic bending of the crystal. In panel (a) the measurement of Wistisen et al. \cite[Fig.8]{WisU16} is shown for the supposed beam entrance angle of $\psi_{0}$ = 0 $\mu$rad, digitized from this figure by the author. It is included for comparison with the simulation calculation also in panel (c). In panel (e) the measurement of Wistisen et al. \cite[Fig.9]{WisU16} is shown for the supposed beam entrance angle of $\psi_{0}$ = 200 $\mu$rad, also digitized by the author. It is included for comparison also in panel (g).} \label{beamProfilesSi6300MeVBentNr1}
\end{figure}

\subsection{Beam profiles for the SLAC experiment at 6.3 GeV} \label{profiles 6300 MeV}
Fig. \ref{beamProfilesSi6300MeVBentNr1} depicts simulated beam profiles and experimental results for 6.3 GeV electrons steered with bent (111) planes of the silicon single crystal. Several interesting features can be recognized. For the spectrum shown in Fig. \ref{beamProfilesSi6300MeVBentNr1} (d) the crystal is  de-tuned already by -60 $\mu$rad towards the amorphous domain. The main peak is situated close at the deflection angle $\Theta$ = 0 mrad, and the re-channeling peak appears shifted to 0.46 mrad, the geometrically expected angle. Its intensity is already rather weak. Increasing the de-tuning angle, the re-channeling peak shifts and reaches the nominal value of $\Theta$ = 0.4 mrad at $\psi_{0}$ = 0 $\mu$rad  as expected. Thereby its intensity increases gradually since more and more electrons will be captured at the crystal entrance into the channel. In panels (f)-(h) simulation calculations for positive de-tuning angles $\psi_{0} > 0$ are depicted. The main peak is shifted towards negative deflection angles  $\Theta$. This is the well known volume reflection.

Let us compare the simulation calculations with the experimental result shown in panel (a) of Fig. \ref{beamProfilesSi6300MeVBentNr1}. The measured beam profile agrees quite well with the simulation calculation shown in panel (c) for $\psi_{0}$ = -30 $\mu$rad, and not for $\psi_{0}$ = 0 $\mu$rad, see black curve which is the experimental result shown in panel (a). The simplest explanation for this finding is that the nominal angular directions in experiment and simulation are differently defined, or in other words, that the crystal is de-tuned with respect to the nominal beam direction. This conjecture is corroborated by means of Fig. \ref{beamProfilesSi6300MeVBentNr1}, panel (g). The measurement agrees best with the simulation at 170 $\mu$rad, and not at 200 $\mu$rad, panel (f), indicating again the de-tuning angle of 30 $\mu$rad. A fine-tuning of $\psi_0$ has not been done.

The intensity of the simulated structure for volume capture appears to be somewhat larger than the measurement indicate, see Fig. \ref{beamProfilesSi6300MeVBentNr1}, panel (g). Also the volume deflection peak is not exactly reproduced by the simulation calculations. It is in comparison to the experiment, black curve, slightly shifted. The reason for these deviations is not quite clear. It may indicate a deficiency of the simulation model. If the particle trajectory approaches the tangent of a bending plane, the effects of the volume-deflection and volume-capture are particularly sensitive to the shape of the electric potential.

The probability density of the deflected particles in Fig. \ref{beamProfilesSi6300MeVBentNr1} (b) for the crystal in the channeling orientation at $\psi_{0}$ = 0 $\mu$rad does not agree with the result obtained with the aid of the DYNECHARM++ simulation by Wienands et al. \cite[Fig. 3]{WieM15}. The channeling peak appears at about 430 and not at 400 $\mu$rad as expected. Apparently the calculations were performed with a de-tuning angle $\psi_{0}$ close to 30 $\mu$rad which was defined as "channeling orientation". Indeed, good agreement is found with our simulation calculation under this assumption.

The DYNECHARM++ simulation result of Wistisen et al. \cite[Fig. 14]{WisU16} is at variance with that of Wienands et al. \cite[Fig. 3]{WieM15} and also our results of Fig. \ref{beamProfilesSi6300MeVBentNr1}.

\section{Discussion} \label{discussion}
For the MAMI experiment at 855 MeV Monte Carlo simulations results are depicted in Mazzolari et al. \cite[Fig. 3]{MazB14} which were made with a code of Baryshevski and Tikhomirov \cite{BarT13}. Nearly perfect agreement with the experimental observation was found for both, alignments of the crystal into the channeling regime ($\psi_0$ = 0), and for the volume capture situation ($\psi_0$ = 450 $\mu$rad). A detailed analysis of the same experiment was performed by Haurylavets et al. \cite{HaL22} an the basis of the MBN explorer software package under various sophisticated assumptions. Results are shown in Fig.  \ref{beamProfilesBent855MeVPsim400murad} suggesting that scattering might be overestimated. The authors mention that a possible reason for the discrepancies can be associated with the Moliére parametrization which describe the electron–atom interaction. Indeed, Fig. \ref{ScatteringFactors} shows that the original Moliére parametrization overestimates for small momentum transfers the scattering factors obtained with the Doyle-Turner approach. This fact influences also the shape of the potential, at least in our continuum potential picture. However, simulation calculations with the original Moliére parameters revealed that this way the differences can not be explained.

For the SLAC experiment at 6.3 GeV \cite{WisU16} Sushko et al. \cite{SusK22} explained the beam profile with the aid of an anticlastic bending of the crystal. Parameter sets for the size of the beam spot, its displacement and the radius of the anticlastic curvature are presented at which the simulations are in accord with the experimental observation.  Their Fig. 2 allows approximately a comparison with our results. The simulation is done for a relatively narrow beam with $\sigma$ = 75 $\mu$m in comparison with the displacement of $h$ = 675 $\mu$m for which a good agreement with the experimental result is obtained. Utilizing their Eqns. (2) and (3) for $\theta_{qm}$ = 400 $\mu$rad and an anticlastic bending radius $R_a$ = 300 cm, the entrance angle is $\theta_e(h)$ = 25 $\mu$rad. The latter is in fair agreement with our finding without the assumption of an anticlastic bending, however, instead for a de-tuning angle of 30 $\mu$rad of the crystal with respect to the beam direction.

\section{Conclusions}\label{conclusions}
In this contribution Monte Carlo simulation results for electron beam deflections in quasi-mosaic bent (111) planes of silicon single crystals were compared with experimental results. The continuum potential picture has been utilized. For the 855 MeV experiment at MAMI good agreement between simulation calculations and the experimental observation was found. For the SLAC experiment at 6.3 GeV the entrance angle into the planes turns out to be a rather sensitive parameter. Small de-tuning angles in the order of some tenth of $\mu$rad's perturb significantly the intensity ratio between transmitted and channeled particles with the consequence, that an anticlastic bending is not necessary to explain the gross properties of the experimental observation. However, it can not be excluded that in an second order approach the remaining small deviations visible in Fig. \ref{beamProfilesSi6300MeVBentNr1}, panels (c) and (g), may require it.

Anyway, in the sense of Occam's razor this simpler explanation should be preferred.


\section*{Acknowledgements} \label{Acknowledgements}
I would like to express my gratitude to José M. Fernández-Varea to provide me with numerical data of H. Bichsel \cite{Bic88} for the optical oscillator strength (OOS) of silicon.

Fruitful discussions with A. V. Korol, W. Lauth, A. V. Solov'yov, A. Mazzolari, and A. Sytov are gratefully acknowledged.

\section*{Declarations}
This work has been financially supported by the European Union’s Horizon 2020 research
and innovation programme – the N-LIGHT project (GA 872196) within the H2020-MSCA-RISE-2019 call.

%
%

\vspace*{1cm}

\begin{appendices}
\section{Differential cross-sections for scattering at electrons} \label{appendix A}
\begin{figure}[b]
\centering
    \includegraphics[angle=0,scale=0.65,clip]{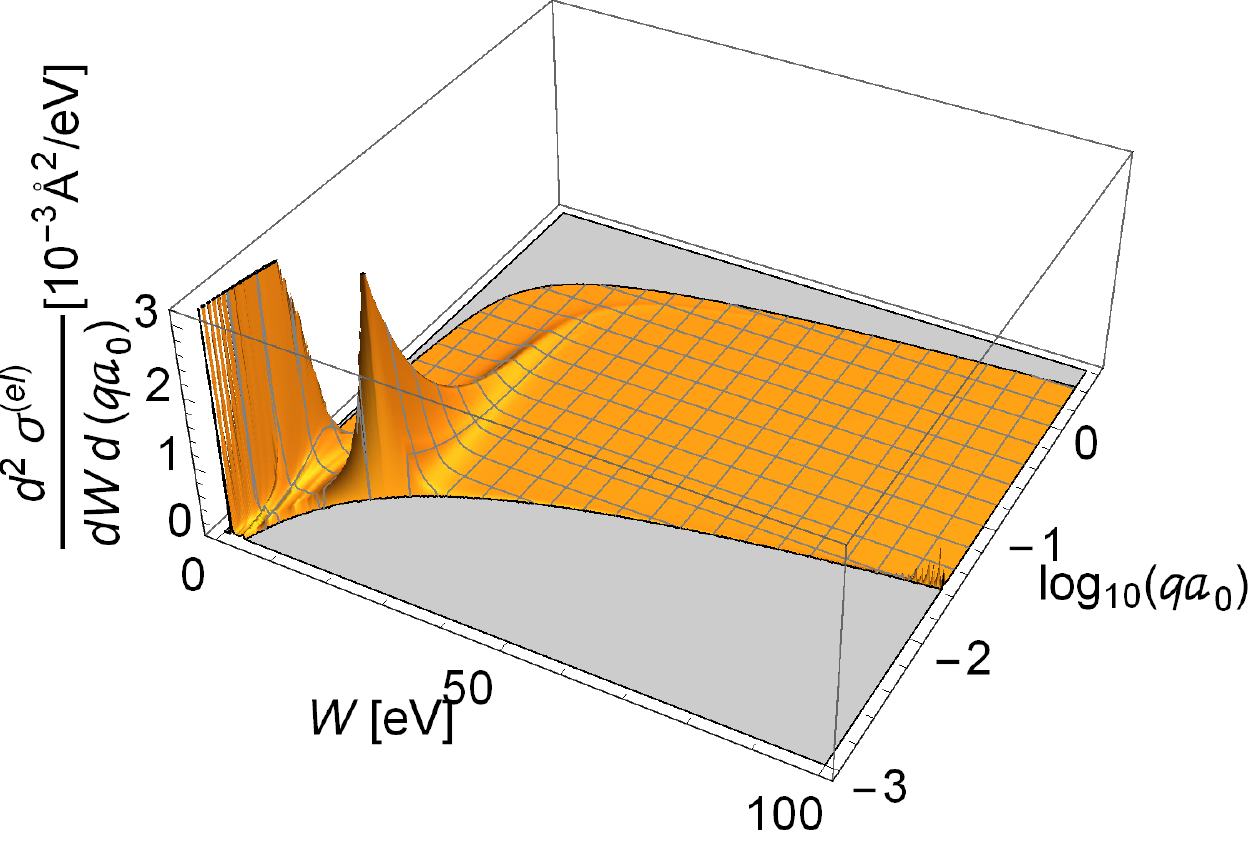}
\caption[]{Double differential cross-section as function of the energy loss $W$ and momentum transfer $q a_0$. The kinematical allowed region lifts clearly out from the gray area. Two dominating features are the plasmon resonance at $W$ = 17 eV and $q a_0 \simeq$ 0.0045 and the resonance of the transverse excitation at $W$ = 3.9 eV and $q a_0 \simeq$  0.0013 which is some orders of magnitudes higher than shown. They have the effect of narrowing the scattering distributions shown in Figs. \ref{ProbabilityWnorm855} and \ref{ProbabilityWnorm6300} for small angles, see \cite{Bac22}. } \label{DoubleDifferentialCrossSection}
\end{figure}

\begin{figure}[tbh]
\centering
\subfigure{
     \includegraphics[angle=0,scale=0.43,clip]{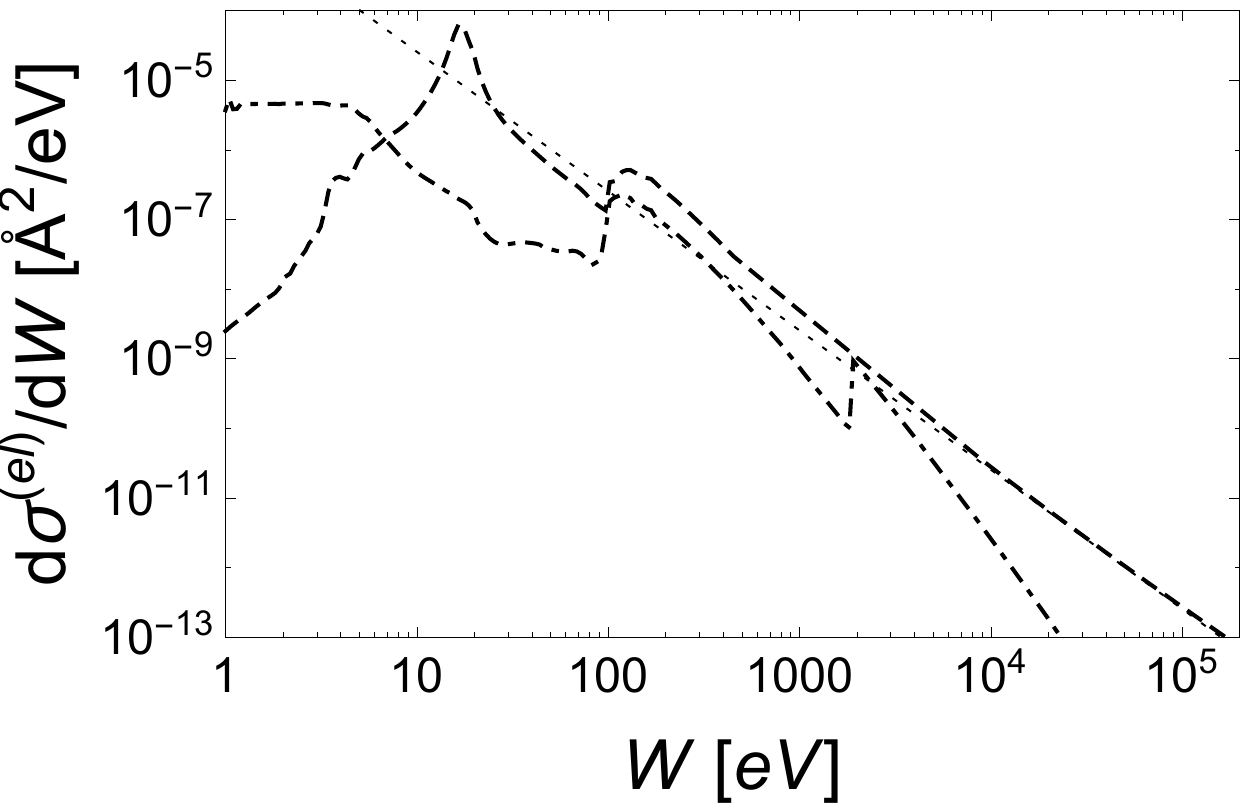}}
    \subfigure{
    \includegraphics[angle=0,scale=0.43,clip]{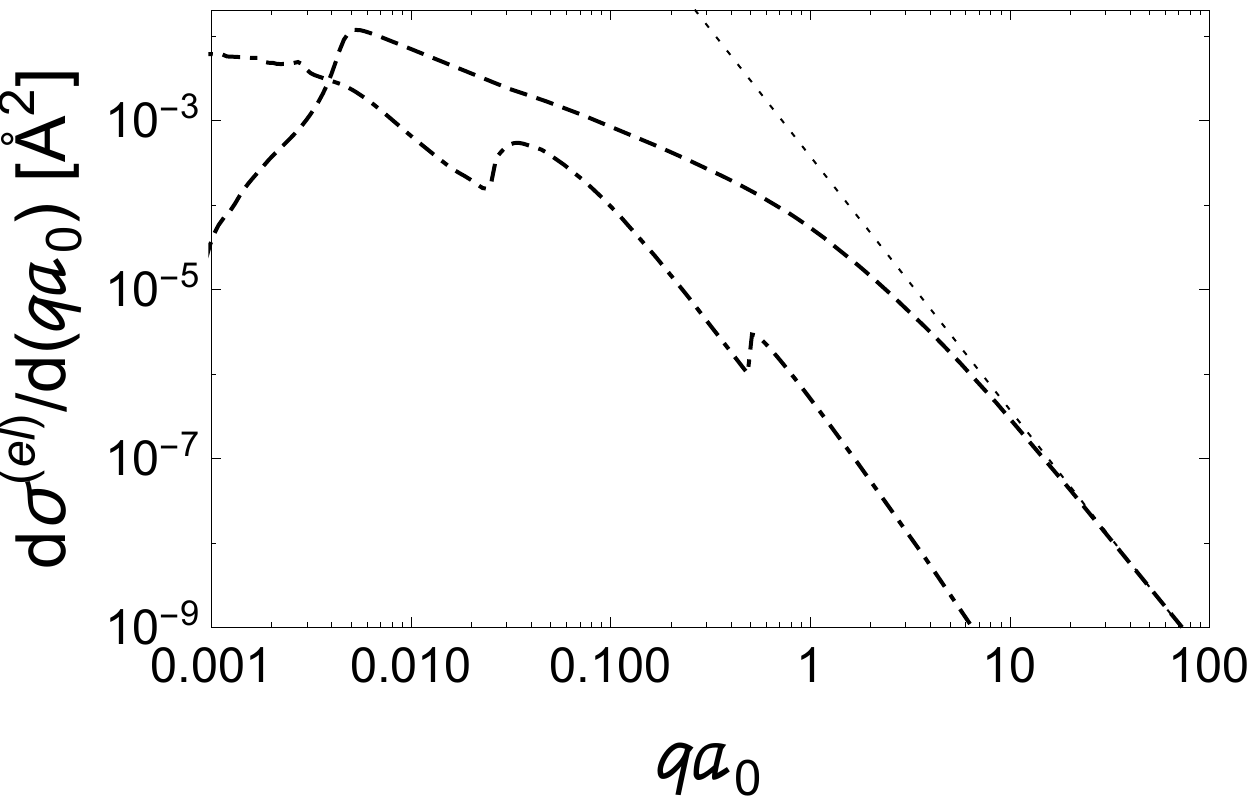}}
\caption[]{Differential cross-sections as function of the energy loss $W$, left panel, and the momentum transfer $q a_0$, right panel. Shown are separately the contributions of the longitudinal (dashed), the transverse (dotted-dashed), and the M{\o}ller cross-sections (dotted).} \label{DifferentialCrossSectionWqa0}
\end{figure}

The low energy part of the double differential cross-section at scattering on atomic electrons is shown in Fig. \ref{DoubleDifferentialCrossSection}.
The energy differential cross-sections as function of the energy loss $W$ and momentum transfer $q a_0$ are obtained after proper integration over the kinematical allowed $q a_0$ and the kinematical allowed $W$, respectively. The results are shown in Fig. \ref{DifferentialCrossSectionWqa0}.
\section{Simulation of de-channeling rates} \label{appendix B}
\begin{figure}[tbh]
     \subfigure{
    {\includegraphics[angle=0,scale=0.57,clip]{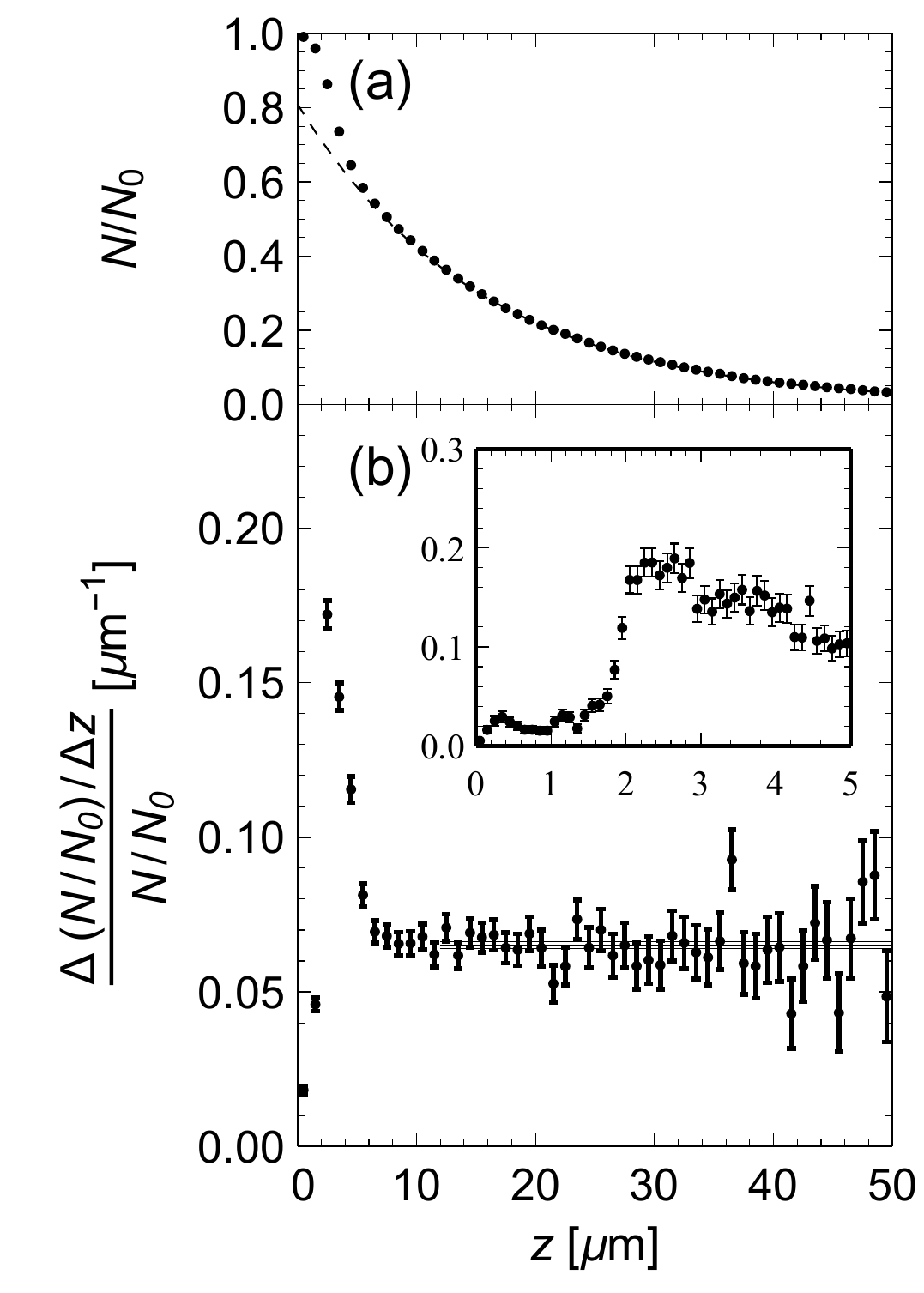}}}
     \subfigure{  \includegraphics[angle=0,scale=0.57,clip]{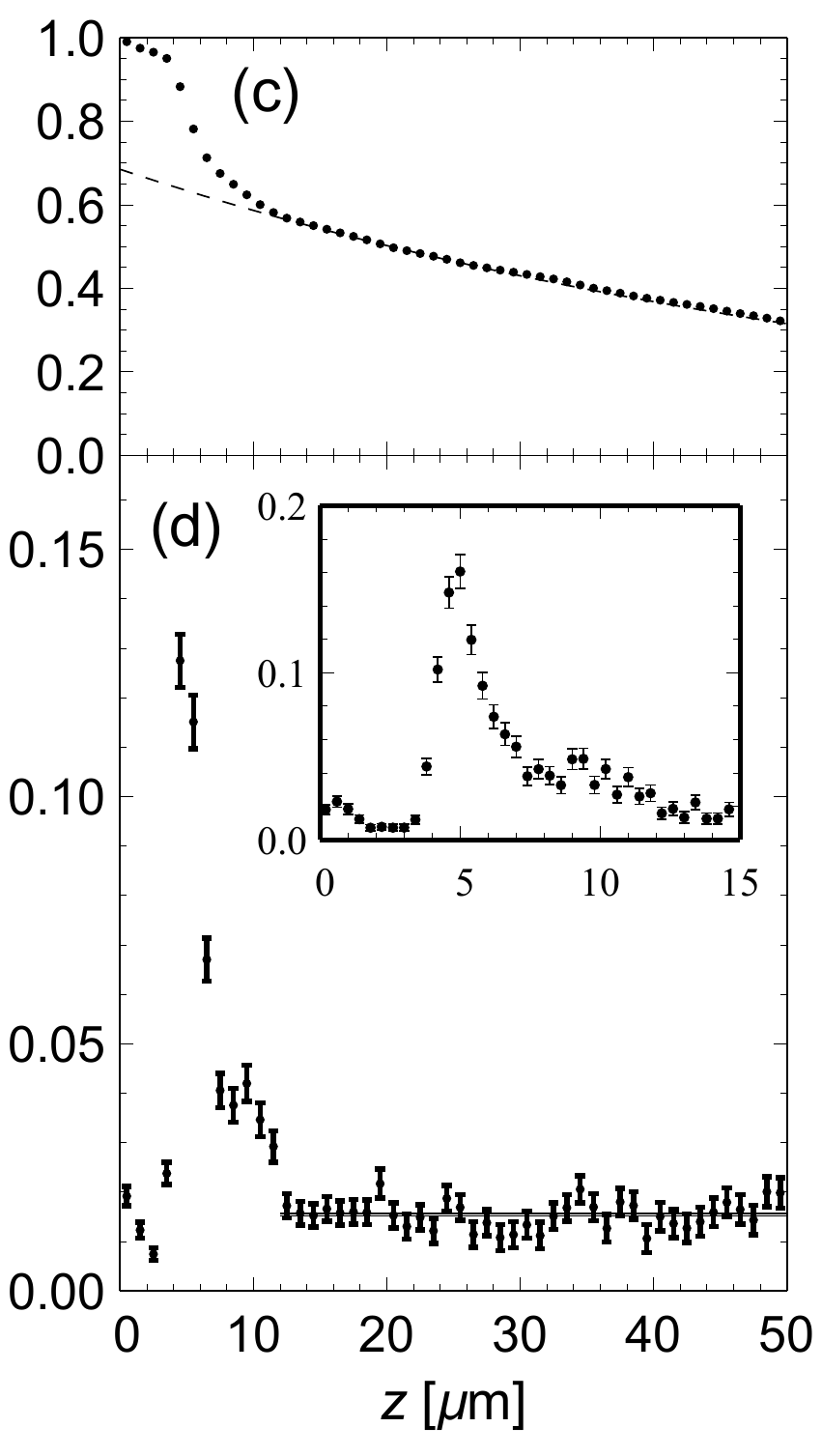}}
\caption[]{Transition rates for (111) channeling in bent silicon single crystals. Panels (a) and (b) for crystal with bending radius 33.5 mm and beam energy 855 MeV. Details for small penetration depths are depicted in the inset. The channel occupation is normalized to unity for $N_0$ = 10,000 events. The horizontal error band indicates the mean value $\overline{\lambda_{de}^{855}(z)}$ = (0.0651 $\pm$ 0.0011)/$\mu$m as obtained for $14 \leq z/\mu m \leq 100$ for 81 \% of particles imping the crystal. Panels (c) and (d) for a crystal with bending radius 150 mm and beam energy of 6.3 GeV. The channel occupation is normalized to unity for $N_0$ = 5,000 events. The horizontal error band indicates the mean value $\overline{\lambda_{de}^{6300}(z)}$ = (0.0155 $\pm$ 0.0003)/$\mu$m as obtained for $14 \leq z/\mu m \leq 150$ for 68.5 \% of particles imping the crystal.} \label{dechannelinRateFirst}
\end{figure}

For the purpose of a check with otherwise obtained results, simulation calculations for the transition rate as function of the penetration depth $z$ were performed for planar (111) channeling of electrons with an energy of 6.3 GeV from which the de-channeling length was derived. De-channeling is signaled if the electron leaves the potential boundaries, the crystal after 200~$\mu$m, or if it reaches the maximum transverse energy for which 20 $U_{0}$ = 452 eV was chosen. Fig. \ref{dechannelinRateFirst}
show results. In panels (b) and (d) instantaneous transition rates
\begin{eqnarray}
\lambda _{de}(z) = -\frac{f_{de}'(z)}{f_{de}(z)} =
\lim_{\substack {\Delta z\to 0 \\ N \to \infty}} \frac{\Delta {(N}(z)/N_0)/\Delta z }{N(z)/N_0},
\label{dechRate}
\end{eqnarray}
are presented as derived from the simulated channel occupation numbers $\Delta {(N}(z)/N_0)/\Delta z$ and the channel occupation $N(z)/N_0$ depicted in panels (a) and (c). For $z$ less than about 14 $\mu$m the instantaneous transition rates are larger than the constant mean value for $z>$ 14 $\mu$m.

For 855 MeV, see Fig. \ref{dechannelinRateFirst} left panel, the simulated de-channeling rate 1/$\overline{\lambda_{de}^{855}(z)}$ = (15.4 $\pm$ 0.3) $\mu$m is in fair agreement with the value 13.6 $\mu$m quoted by Mazzolari et al. \cite[Table I]{MazB14}. However, the latter is not based on pure experimental observations but relies also on simulation calculations. For 6.3 GeV, see Fig. \ref{dechannelinRateFirst} right panel, the simulated de-channeling length 1/$\overline{\lambda_{de}^{6300}(z)}$ = (64.4 $\pm$ 1.3) $\mu$m agrees well with the value (65.3 $\pm$ 1.9) $\mu$m quoted by Wistinsen et al. \cite[Table IV]{WisU16}. It should be mentioned that the value originally reported by Wienands et al. \cite{WieM15} is siginificantly lower. The simulated channeling efficiency of 68.5 \%, also called surface transmission, is somewhat larger than the experimental result, however, in accord with DYNECHARM++ simulations \cite[Table V]{WisU16}.


\end{appendices}

\newpage


\bibliography{bibfileBa}


\begin{thebibliography}{22}
\ifx \bisbn   \undefined \def \bisbn  #1{ISBN #1}\fi
\ifx \binits  \undefined \def \binits#1{#1}\fi
\ifx \bauthor  \undefined \def \bauthor#1{#1}\fi
\ifx \batitle  \undefined \def \batitle#1{#1}\fi
\ifx \bjtitle  \undefined \def \bjtitle#1{#1}\fi
\ifx \bvolume  \undefined \def \bvolume#1{\textbf{#1}}\fi
\ifx \byear  \undefined \def \byear#1{#1}\fi
\ifx \bissue  \undefined \def \bissue#1{#1}\fi
\ifx \bfpage  \undefined \def \bfpage#1{#1}\fi
\ifx \blpage  \undefined \def \blpage #1{#1}\fi
\ifx \burl  \undefined \def \burl#1{\textsf{#1}}\fi
\ifx \doiurl  \undefined \def \doiurl#1{\url{https://doi.org/#1}}\fi
\ifx \betal  \undefined \def \betal{\textit{et al.}}\fi
\ifx \binstitute  \undefined \def \binstitute#1{#1}\fi
\ifx \binstitutionaled  \undefined \def \binstitutionaled#1{#1}\fi
\ifx \bctitle  \undefined \def \bctitle#1{#1}\fi
\ifx \beditor  \undefined \def \beditor#1{#1}\fi
\ifx \bpublisher  \undefined \def \bpublisher#1{#1}\fi
\ifx \bbtitle  \undefined \def \bbtitle#1{#1}\fi
\ifx \bedition  \undefined \def \bedition#1{#1}\fi
\ifx \bseriesno  \undefined \def \bseriesno#1{#1}\fi
\ifx \blocation  \undefined \def \blocation#1{#1}\fi
\ifx \bsertitle  \undefined \def \bsertitle#1{#1}\fi
\ifx \bsnm \undefined \def \bsnm#1{#1}\fi
\ifx \bsuffix \undefined \def \bsuffix#1{#1}\fi
\ifx \bparticle \undefined \def \bparticle#1{#1}\fi
\ifx \barticle \undefined \def \barticle#1{#1}\fi
\bibcommenthead
\ifx \bconfdate \undefined \def \bconfdate #1{#1}\fi
\ifx \botherref \undefined \def \botherref #1{#1}\fi
\ifx \url \undefined \def \url#1{\textsf{#1}}\fi
\ifx \bchapter \undefined \def \bchapter#1{#1}\fi
\ifx \bbook \undefined \def \bbook#1{#1}\fi
\ifx \bcomment \undefined \def \bcomment#1{#1}\fi
\ifx \oauthor \undefined \def \oauthor#1{#1}\fi
\ifx \citeauthoryear \undefined \def \citeauthoryear#1{#1}\fi
\ifx \endbibitem  \undefined \def \endbibitem {}\fi
\ifx \bconflocation  \undefined \def \bconflocation#1{#1}\fi
\ifx \arxivurl  \undefined \def \arxivurl#1{\textsf{#1}}\fi
\csname PreBibitemsHook\endcsname

\bibitem{KorS14}
\begin{bbook}
\bauthor{\bsnm{Korol}, \binits{A.V.}},
\bauthor{\bsnm{Solov'yov}, \binits{A.V.}},
\bauthor{\bsnm{Greiner}, \binits{W.}}:
\bbtitle{{Channeling and Radiation in Periodically Bent Crystals, 2nd Edn.}}
\bpublisher{Springer},
\blocation{{Springer-Verlag Berlin Heidelberg 2014}}
(\byear{2014}).
\bcomment{{Springer Series on Atomic, Optical and Plasma Physics 69, 2014}}
\end{bbook}
\endbibitem

\bibitem{CamG15}
\begin{barticle}
\bauthor{\bsnm{Camattari}, \binits{R.}},
\bauthor{\bsnm{Guidi}, \binits{V.}},
\bauthor{\bsnm{Bellucci}, \binits{V.}},
\bauthor{\bsnm{Mazzolari}, \binits{A.}}:
\batitle{{The ‘quasi-mosaic’ effect in crystals and its applications in
  modern physics}}.
\bjtitle{Journal of Applied Crystallography}
\bvolume{48},
\bfpage{977}--\blpage{989}
(\byear{2015}).
\doiurl{0.1107/S1600576715009875}.
\bcomment{{}}
\end{barticle}
\endbibitem

\bibitem{MazB14}
\begin{barticle}
\bauthor{\bsnm{Mazzolari}, \binits{A.}},
\bauthor{\bsnm{Bagli}, \binits{E.}},
\bauthor{\bsnm{Bandiera}, \binits{L.}},
\bauthor{\bsnm{Guidi}, \binits{V.}},
\bauthor{\bsnm{Backe}, \binits{H.}},
\bauthor{\bsnm{Lauth}, \binits{W.}},
\bauthor{\bsnm{Tikhomirov}, \binits{V.}},
\bauthor{\bsnm{Berra}, \binits{A.}},
\bauthor{\bsnm{Lietti}, \binits{D.}},
\bauthor{\bsnm{Prest}, \binits{M.}},
\bauthor{\bsnm{Vallazza}, \binits{E.}},
\bauthor{\bsnm{De~Salvador}, \binits{D.}}:
\batitle{{Steering of a Sub-GeV Electron Beam through Planar Channeling
  Enhanced by Rechanneling}}.
\bjtitle{Physical Review Letters}
\bvolume{112},
\bfpage{135503}
(\byear{2014}).
\doiurl{10.1103/PhysRevLett.112.135503}
\end{barticle}
\endbibitem

\bibitem{WieM15}
\begin{barticle}
\bauthor{\bsnm{Wienands}, \binits{U.}},
\bauthor{\bsnm{Markiewicz}, \binits{T.W.}},
\bauthor{\bsnm{Nelson}, \binits{J.}},
\bauthor{\bsnm{Noble}, \binits{R.J.}},
\bauthor{\bsnm{Turner}, \binits{J.L.}},
\bauthor{\bsnm{Uggerh{\o}j}, \binits{U.I.}},
\bauthor{\bsnm{Wistisen}, \binits{T.N.}},
\bauthor{\bsnm{Bagli}, \binits{E.}},
\bauthor{\bsnm{Bandiera}, \binits{L.}},
\bauthor{\bsnm{Germogli}, \binits{G.}},
\bauthor{\bsnm{Guidi}, \binits{V.}},
\bauthor{\bsnm{Mazzolari}, \binits{A.}},
\bauthor{\bsnm{Holtzapple}, \binits{R.}},
\bauthor{\bsnm{Miller}, \binits{M.}}:
\batitle{{Observation of Deflection of a Beam of Multi-GeV Electrons by a Thin
  Crystal}}.
\bjtitle{Physical Review Letters}
\bvolume{114},
\bfpage{074801}--\blpage{6}
(\byear{2015}).
\doiurl{10.1103/PhysRevLett.114.074801}
\end{barticle}
\endbibitem

\bibitem{WisU16}
\begin{barticle}
\bauthor{\bsnm{Wistisen}, \binits{T.N.}},
\bauthor{\bsnm{Uggerh{\o}j}, \binits{U.I.}},
\bauthor{\bsnm{Wienands}, \binits{U.}},
\bauthor{\bsnm{Markiewicz}, \binits{T.W.}},
\bauthor{\bsnm{Noble}, \binits{R.J.}},
\bauthor{\bsnm{Benson}, \binits{B.C.}},
\bauthor{\bsnm{Smith}, \binits{T.}},
\bauthor{\bsnm{Bagli}, \binits{E.}},
\bauthor{\bsnm{Bandiera}, \binits{L.}},
\bauthor{\bsnm{Germogli}, \binits{G.}},
\bauthor{\bsnm{Guidi}, \binits{V.}},
\bauthor{\bsnm{Mazzolari}, \binits{A.}},
\bauthor{\bsnm{Holtzapple}, \binits{R.}},
\bauthor{\bsnm{Tucker}, \binits{S.}}:
\batitle{{Channeling, volume reflection, and volume capture study of electrons
  in a bent silicon crystal}}.
\bjtitle{Physical Review Accelerators and Beams}
\bvolume{19},
\bfpage{071001}--\blpage{11}
(\byear{2016}).
\doiurl{10.1103/PhysRevAccelBeams.19.071001}
\end{barticle}
\endbibitem

\bibitem{GuiM09}
\begin{barticle}
\bauthor{\bsnm{Guidi}, \binits{V.}},
\bauthor{\bsnm{Mazzolari}, \binits{A.}},
\bauthor{\bsnm{Salvador}, \binits{D.D.}},
\bauthor{\bsnm{Carnera}, \binits{A.}}:
\batitle{{Silicon crystal for channelling of negatively charged particles}}.
\bjtitle{Journal of Physics D: Applied Physics}
\bvolume{42},
\bfpage{182005}
(\byear{2009})
\end{barticle}
\endbibitem

\bibitem{SusK21}
\begin{botherref}
\oauthor{\bsnm{Sushko}, \binits{G.B.}},
\oauthor{\bsnm{Korol}, \binits{A.V.}},
\oauthor{\bsnm{Solov'yov}, \binits{A.V.}}:
{Ultra-relativistic electron beams deflection by quasi-mosaic crystals}.
{arXiv:submit/3994928 [physics.acc-ph]},
1--7
(2021)
\end{botherref}
\endbibitem

\bibitem{Lin65}
\begin{barticle}
\bauthor{\bsnm{Lindhard}, \binits{J.}}:
\batitle{{Influence of Crystal Lattice on Motion of Energetic Charged
  Particles}}.
\bjtitle{Mat. Fys. Medd. Dan. Vid. Selsk.}
\bvolume{34 no.14},
\bfpage{1}--\blpage{64}
(\byear{1965}).
\bcomment{{}}
\end{barticle}
\endbibitem

\bibitem{KorS21}
\begin{barticle}
\bauthor{\bsnm{Korol}, \binits{A.V.}},
\bauthor{\bsnm{Sushko}, \binits{G.B.}},
\bauthor{\bsnm{Solov'yov}, \binits{A.V.}}:
\batitle{{All-atom relativistic molecular dynamics simulations of channeling
  and radiation processes in oriented crystals}}.
\bjtitle{The European Physical Journal D}
\bvolume{75},
\bfpage{107}
(\byear{2021}).
\doiurl{10.1140/epjd/s10053-021-00111-w}.
\bcomment{{}}
\end{barticle}
\endbibitem

\bibitem{BagG13}
\begin{barticle}
\bauthor{\bsnm{Bagli}, \binits{E.}},
\bauthor{\bsnm{Guidi}, \binits{V.}}:
\batitle{{DYNECHARM++: a toolkit to simulate coherent interactions of
  high-energy charged particles in complex structures}}.
\bjtitle{Nuclear Instruments and Methods in Physics Research B}
\bvolume{309},
\bfpage{124}--\blpage{129}
(\byear{2013}).
\doiurl{10.1016/j.nimb.2013.01.073}.
\bcomment{{}}
\end{barticle}
\endbibitem

\bibitem{SytT19}
\begin{barticle}
\bauthor{\bsnm{Sytov}, \binits{A.I.}},
\bauthor{\bsnm{Tikhomirov}, \binits{V.V.}},
\bauthor{\bsnm{Bandiera}, \binits{L.}}:
\batitle{{Simulation code for modeling of coherent effects of radiation
  generation in oriented crystals}}.
\bjtitle{{Physical Review Accelerators and Beams}}
\bvolume{22},
\bfpage{064601}--\blpage{10}
(\byear{2019}).
\doiurl{10.1103/PhysRevAccelBeams.22.064601}
\end{barticle}
\endbibitem

\bibitem{Bac22}
\begin{botherref}
\oauthor{\bsnm{Backe}, \binits{H.}}:
{De-channeling in terms of instantaneous transition rates - Computer
  simulations for 855 MeV electrons at (110) planes of diamond}.
{arXiv:submit/4273372 [physics.data-an]},
1--14
(2022)
\end{botherref}
\endbibitem

\bibitem{Mol47}
\begin{barticle}
\bauthor{\bsnm{Moli\`{e}re}, \binits{G.}}:
\batitle{{Theorie der Streuung schneller geladener Teilchen I, Einzelstreuung
  am abgeschirmten Coulomb-Feld}}.
\bjtitle{Zeitschrift f\"{u}r Naturforschung}
\bvolume{2 a},
\bfpage{133}--\blpage{145}
(\byear{1947})
\end{barticle}
\endbibitem

\bibitem{ChoU99}
\begin{barticle}
\bauthor{\bsnm{Chouffani}, \binits{K.}},
\bauthor{\bsnm{\"{U}berall}, \binits{H.}}:
\batitle{{Theory of Low Energy Channeling Radiation: Application to a Germanium
  Crystal}}.
\bjtitle{Physica Status Solidi (b)}
\bvolume{213},
\bfpage{107}--\blpage{151}
(\byear{1999}).
\bcomment{{}}
\end{barticle}
\endbibitem

\bibitem{DoyT67}
\begin{barticle}
\bauthor{\bsnm{Doyle}, \binits{P.A.}},
\bauthor{\bsnm{Turner}, \binits{P.S.}}:
\batitle{{Relativistic Hartree-Fock X-ray and Electron Scattering Factors}}.
\bjtitle{Acta Crystallographica A}
\bvolume{24},
\bfpage{390}--\blpage{397}
(\byear{1968}).
\bcomment{{}}
\end{barticle}
\endbibitem

\bibitem{website:ioffe}
\begin{botherref}
\url{\\http://www.ioffe.ru/SVA/NSM/Semicond/Diamond/basic.html}
\end{botherref}
\endbibitem

\bibitem{BaiK98}
\begin{bbook}
\bauthor{\bsnm{Baier}, \binits{V.N.}},
\bauthor{\bsnm{Katkov}, \binits{V.M.}},
\bauthor{\bsnm{Strakhovenko}, \binits{V.M.}}:
\bbtitle{{Electromagnetic Processes at High Energies in Oriented Single
  Crystals}}.
\bpublisher{World Scientific, Singapore, New Jersey, London, HongKong},
\blocation{{World Scientific Publishing Co. Pte. Ltd, P O Box 128, Farrer Road,
  Singapore 912805}}
(\byear{1998}).
\bcomment{{}}
\end{bbook}
\endbibitem

\bibitem{BacL08}
\begin{barticle}
\bauthor{\bsnm{Backe}, \binits{H.}},
\bauthor{\bsnm{Kunz}, \binits{P.}},
\bauthor{\bsnm{Lauth}, \binits{W.}},
\bauthor{\bsnm{Rueda}, \binits{A.}}:
\batitle{{Planar channeling experiments with electrons at the 855 MeV Mainz
  Microtron MAMI}}.
\bjtitle{Nuclear Instruments and Methods in Physics Research B}
\bvolume{266}(\bissue{17}),
\bfpage{3835}--\blpage{3851}
(\byear{2008}).
\doiurl{10.1016/j.nimb.2008.05.012}.
\bcomment{Radiation from Relativistic Electrons in Periodic Structures
  RREPS’07}
\end{barticle}
\endbibitem

\bibitem{HaL22}
\begin{barticle}
\bauthor{\bsnm{Haurylavets}, \binits{V.V.}},
\bauthor{\bsnm{Leukovich}, \binits{A.}},
\bauthor{\bsnm{Sytov}, \binits{A.}},
\bauthor{\bsnm{Bandiera}, \binits{L.}},
\bauthor{\bsnm{Mazzolari}, \binits{A.}},
\bauthor{\bsnm{Romagnoni}, \binits{M.}},
\bauthor{\bsnm{Guidi}, \binits{V.}},
\bauthor{\bsnm{Sushko}, \binits{G.B.}},
\bauthor{\bsnm{Korol}, \binits{A.V.}},
\bauthor{\bsnm{Solov’yov}, \binits{A.V.}}:
\batitle{{MBN explorer atomistic simulations of 855 MeV electron propagation
  and radiation emission in oriented silicon bent crystal: theory versus
  experiment}}.
\bjtitle{The European Physical Journal Plus}
\bvolume{137},
\bfpage{137}--\blpage{34}
(\byear{2022}).
\doiurl{10.1140/epjp/s13360-021-02268-0}.
\bcomment{{}}
\end{barticle}
\endbibitem

\bibitem{BarT13}
\begin{barticle}
\bauthor{\bsnm{Baryshevsky}, \binits{V.G.}},
\bauthor{\bsnm{Tikhomirov}, \binits{V.V.}}:
\batitle{{Crystal undulators: from the prediction to the mature simulations}}.
\bjtitle{Nuclear Instruments and Methods in Physics Research B}
\bvolume{309},
\bfpage{30}--\blpage{36}
(\byear{2013}).
\bcomment{{}}
\end{barticle}
\endbibitem

\bibitem{SusK22}
\begin{botherref}
\oauthor{\bsnm{Sushko}, \binits{G.B.}},
\oauthor{\bsnm{Korol}, \binits{A.V.}},
\oauthor{\bsnm{Solov'yov}, \binits{A.V.}}:
{Ultra-relativistic electron beams deflection by quasi-mosaic crystals}.
{arXiv:submit/2110.12959v2 [physics.acc-ph]},
1--6
(2022)
\end{botherref}
\endbibitem

\bibitem{Bic88}
\begin{barticle}
\bauthor{\bsnm{Bichsel}, \binits{H.}}:
\batitle{{Straggling in thin silicon detectors}}.
\bjtitle{Reviews of Modern Physics}
\bvolume{3},
\bfpage{663}--\blpage{699}
(\byear{1988}).
\bcomment{{}}
\end{barticle}
\endbibitem

\end{thebibliography}


\end{document}